\documentclass[aps,prc,twocolumn,superscriptaddress,showpacs,nofootinbib,floatfix]{revtex4}
\usepackage{graphicx}
\usepackage{dcolumn}
\usepackage{bm}
\usepackage{epsfig}

\def\Journal#1#2#3#4{{#1}{\bf #2}, #3 (#4)}

\def\EPJ{{Eur. Phys. J.}~{\bf C}~}

\def\NPA{{Nucl. Phys.}~{\bf A}}
\def\NPB{{Nucl. Phys.}~{\bf B}}
\def\PLB{{Phys. Lett.}~{\bf B}~}
\def\PR{Phys. Repts.~}
\def\PRL{Phys. Rev. Lett.~}
\def\PRL{Phys. Rev.~}
\def\PRD{{Phys. Rev.}~{\bf D}~}
\def\PRC{{Phys. Rev.}~{\bf C}~}

\begin{document}
\title{Medium induced jet absorption in relativistic heavy ion collisions}
\newcommand{\stonybrook}{Department of Physics and Astronomy, State University of New York, Stony Brook, NY 11794-3800}
\newcommand{\columbia}{Current Address: Columbia University, New York, NY 10027 and Nevis Laboratories, Irvington, NY 10533, USA}
\affiliation{\stonybrook}\affiliation{\columbia}
\author{Axel~Drees}\affiliation{\stonybrook}
\author{Haidong~Feng}\affiliation{\stonybrook}
\author{Jiangyong~Jia}\affiliation{\stonybrook}\affiliation{\columbia}
\date{\today}
\begin{abstract}
The dense medium created in Au + Au collisions at the Relativistic
Heavy-Ion Collider (RHIC) significantly suppresses particle
production from hard scattering processes and their characteristic
back-to-back angular correlation. We present a simple model of jet
absorption in dense matter which incorporates a realistic nuclear
geometry. Our calculations are performed at the jet level and
assume independent jet fragmentation in the vacuum. This model
describes quantitatively the centrality dependence of the observed
suppression of the high $p_T$ hadron yield and of the back-to-back
angular correlations. The azimuthal anisotropy of high $p_T$
particle production can not be accounted for using a realistic
nuclear geometry.
\end{abstract}
 \maketitle

\section{Introduction}
In relativistic heavy ion collisions at 62.4 $< \sqrt{s_{NN}} <$ 200 GeV in the
Relativistic Heavy-Ion Collider (RHIC), partons scattered with
large $Q^2$ become a valuable tool to study nuclear matter under
extreme conditions. Due to the large $Q^2$, the hard scattering
occurs early in the collisions, thus the scattered partons may
directly probe the subsequently produced hot, dense and strongly
interacting medium.

In collisions of elementary particles, i.e. in the absence of a
dense medium, the hard scattered partons typically fragment into
two back-to-back jets of hadrons. These hadrons have large
transverse momentum ($p_T$) and pronounced azimuthal angular
correlations. The production cross sections can be calculated with
perturbative QCD (pQCD) based on the fragmentation theorem and
extrapolated to heavy ion collisions. Theoretical studies predict
that the hard scattered partons suffer a large energy loss by
multiple scattering and induced gluon radiation as they propagate
through dense matter~\cite{Gyu90}. Unlike energy loss in QED, the
radiative gluon energy loss per unit length $dE/dx$ not only
depends on the color charge density and the momentum distribution
of the partons in the medium, but also linearly depends on the
thickness of the medium, due to the non-Abelian nature of gluon
radiation in QCD~\cite{BDMPS,baier}.

Data from Au + Au collisions at $\sqrt{s_{NN}}$ = 130 and 200 GeV have
revealed rich information on high $p_T$ phenomena. One of the most
interesting results is the apparent ``jet quenching'' in central
collisions, observed as suppression of the hadron yield by a
factor of 4--5~\cite{ppg014,ppg023,starch} and hadron back-to-back
correlation strength by a factor of 5--10~\cite{starbtob},
compared to expectations based on the underlying nucleon-nucleon
collisions. The absence of these phenomena in $d$ + Au collisions
suggests that the observed suppression in central Au + Au
collisions is indeed an effect of the dense medium created during
the collisions, consistent with parton energy loss in the dense
medium~\cite{ppg028,daustar,dauphobos,daubrahms}.

Both the suppression of the high $p_T$ hadron yield and the
back-to-back angular correlations show a characteristic centrality
dependence, which seems to be independent of $p_T$ for $p_T~>$~4.5
GeV/$c$~\cite{ppg014,ppg023,starch}. Interestingly, in this $p_T$
range, particle production seems consistent with jet
fragmentation, despite the suppression. Specifically, experiments
have observed:
\begin{enumerate}
\item
An identical spectral shape compared to $p$ + $p$ collisions within
systematic errors~\cite{ppg023,starch}.
\item
A similar suppression for charged hadrons and
$\pi^0$'s~\cite{ppg023} and for charged hadron, $\Lambda$, and
$K^0_s$~\cite{starks}.
\item
A $h/\pi^0$ ratio consistent with values observed in $p$ + $p$
collisions, indicating a similar particle composition in $p$ + $p$ and
Au + Au at high $p_T$~\cite{ppg015,ppg023}.
\item
Strength, width, and charge composition of near angle correlation
consistent with jet fragmentation~\cite{starbtob}.
\item
$x_T$ scaling of pion production cross section in Au + Au between
$\sqrt{s_{NN}}$ = 130 and 200 GeV is similar to $p$ + $p$
collisions~\cite{ppg023}.
\end{enumerate}
The data suggest that high $p_T$ particles come dominantly from
jet fragmentation in vacuum.

In relativistic heavy-ion collisions, energy is deposited in the
overlap region between two colliding nuclei. The size, shape and
energy density of the medium formed in this region strongly
depends on the impact parameter of the collision. The amount of
medium a hard scattered parton traverses, and subsequently its
energy loss, varies with the centrality of the collision and also
the azimuthal angle with respect to the reaction plane. If the
parton energy loss is large, the surviving partons will be emitted
dominantly near the outer layer of the overlap
region~\cite{bjorken2}. The partons moving towards the surface
(near side) traverse on average less material than those going in
opposite direction (away side). Thus partons scattered to the near
side are likely to escape with little energy loss, while the away
side partons are likely to lose significant energy and thus are
suppressed more strongly.

It was shown before~\cite{xinnianprl,ivanprl,ursprl,baier2} that
for a static non-Abelian partonic medium, the mean total energy
loss depends quadratically on the medium size. In addition, the
`dynamical scaling law` allows to write the parton energy loss for
an expanding medium in terms of an equivalent static medium with a
linear dependence on the path length~\cite{ursprl}. Thus to the 
first order, one can decouple details of the energy loss
formulation, which is complicated and model dependent, from the
simple geometry of the medium. Recently, several authors have
modelled the energy loss of hard scattered partons in Au + Au
collisions ~\cite{xnwang3,vitevv2,mueller,nara,wangnew,shuryak2}.
Different theoretical approaches for the energy loss have been
used and compared to experimental data like the nuclear
modification factor $R_{AA}$, the back-to-back jet correlation
strength, and the azimuthal anisotropy of particle production
$v_2$. In general, reasonable agreement between the model
calculations and experimental data can be achieved.

In this paper, we present a simple model of jet absorption in an
extremely opaque medium, following the approach in
Ref.~\cite{shuryak1} that was used to set a limit on the expected
azimuthal anisotropy due to jet quenching. The model is equivalent
to an extreme jet energy loss scenario in which most of the energy
is lost in a single scattering. We use this simplified model to
focus on the geometric aspect of the jet propagation in dense
medium. After incorporating a realistic nuclear geometry, we
demonstrate that the centrality dependence of the high $p_T$
suppression of the yield and of the back-to-back correlation
strength can be described quantitatively. We also discuss the
sensitivity of our results on the jet absorption patterns and
different collision geometry assumptions, and make predictions for
the system size dependence.

\begin{figure*}[t]
\begin{center}
\begin{tabular}{c}
\epsfig{file=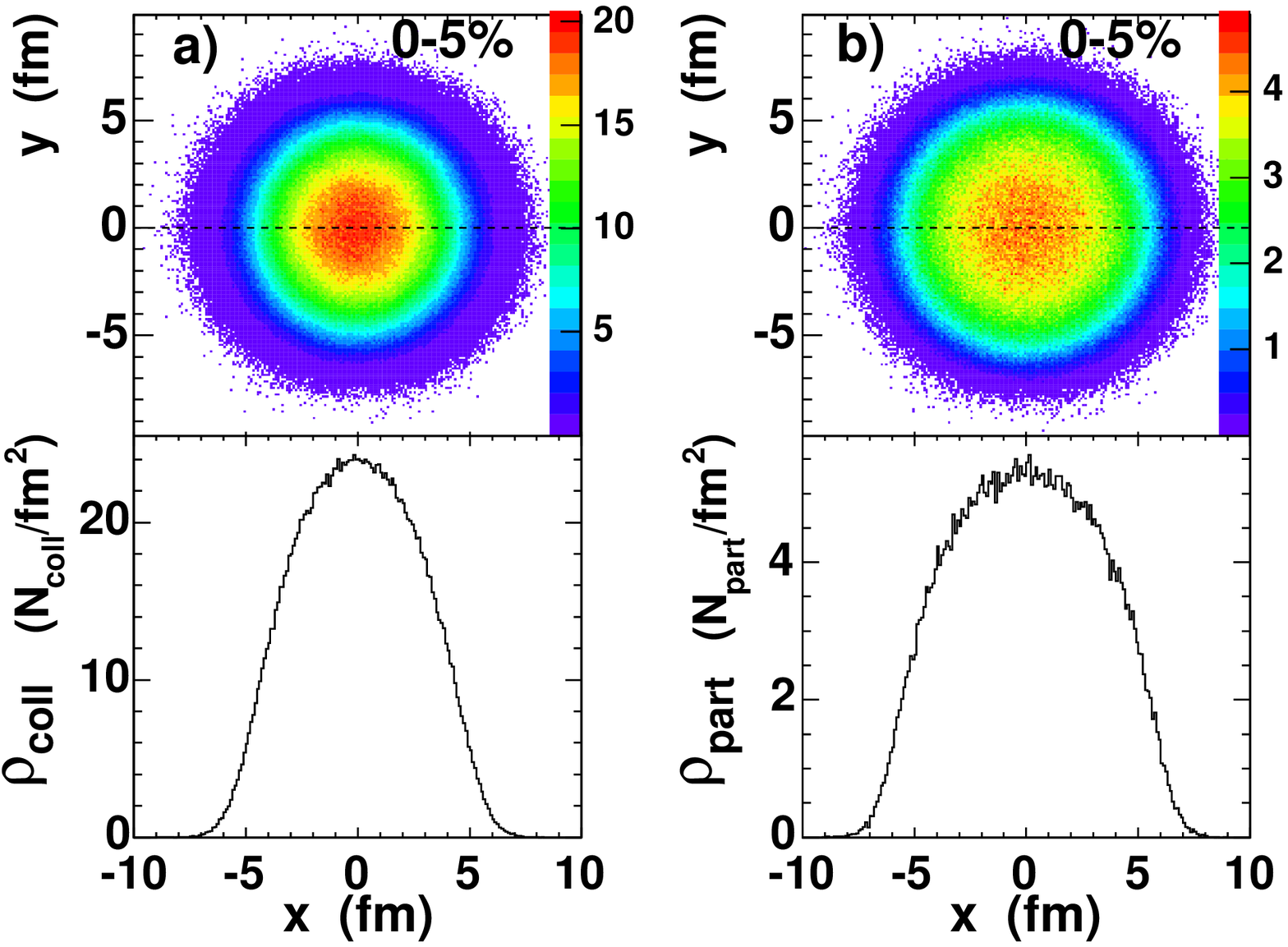,width=3.4in}\hspace*{0in}
\epsfig{file=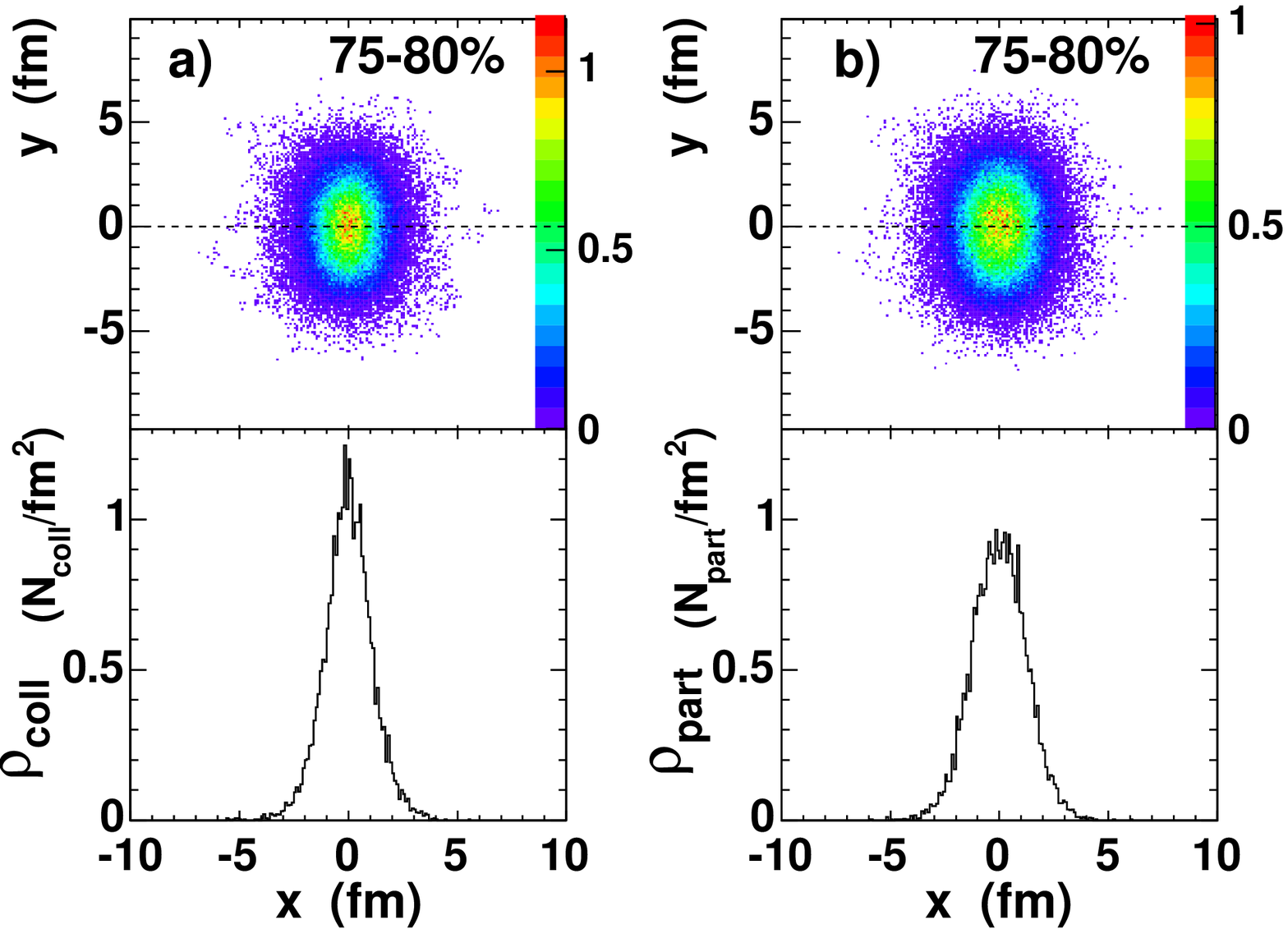,width=3.4in}
\end{tabular}
\caption{\label{fig:mden} (Color online) Collision density
$\rho_{\textrm{coll}}(x,y)$ (a) and participant density $\rho_{\textrm{part}}(x,y)$
(b) in transverse plane for central (0--5\%) and peripheral
(75--80\%) collisions. The bottom panels are the average density
along the impact parameter axis, which is indicated by the dashed
line in the top panels.}
\end{center}
\end{figure*}

\begin{table*}
\caption{\label{tab:model1}. Centrality dependence of parameters
calculated with a Monte Carlo based Glauber approach. For each
centrality class, we list the average number of participants, the
average number of nucleon-nucleon collisions, the average impact
parameter, the average participant density in the overlap region,
the maximum participant density, the maximum nucleon-nucleon
collision density, the maximum energy density calculated using the
Bjorken estimate at $\tau_0 = 1~\textrm{fm}/c$(see text), and the transverse area with a
density larger than 1 GeV/$\textrm{fm}^3$.}
\begin{ruledtabular} \begin{tabular}{lllllllll}
Centrality &$\langle N_{\textrm{part}} \rangle$ &$\langle N_{\textrm{coll}}\rangle$ &
$\left<b\right>$ $\left(\textrm{fm}\right)$&$\left<\rho_{\textrm{part}}\right>$ $\left(\textrm{fm}^{-2}\right)$&
$\rho^{\textrm{max}}_{\textrm{part}}$ $\left(\textrm{fm}^{-2}\right)$&$\rho^{\textrm{max}}_{\textrm{coll}}$ $\left(\textrm{fm}^{-2}\right)$&
$\epsilon^{\textrm{max}}_{\textrm{bj}} \left(\textrm{GeV/fm}^3\right)$&$A\left(\epsilon_{\textrm{bj}}~>~1\right)$ $\left(\textrm{fm}^2\right)$\\
&          &                                    &                                                             &
                                                &                                                                   &
                                                                     & $\left(\tau_0~=~1.0~\textrm{fm}/c\right)$&
\\\hline
0--5\%  &350  &1090 &2.2  &2.48  &4.2  &18.9 &8.5  & 138\\
15--20\%&215  &540  &6.2  &2.10  &3.8  &15.0 &7.6  & 94.7\\
30--35\%&125  &250  &8.5  &1.76  &3.1  &10.3 &6.3  & 64.7\\
45--50\%&67   &105  &10.2 &1.42  &2.4  &6.2  &4.9  & 41.6\\
60--65\%&30   &35   &11.7 &1.09  &1.5  &2.7  &3.1  & 23.2\\
75--80\%&11   &10   &13   &0.79  &0.76 &0.9  &1.5   & 6.9\\
90--95\%&4.1  &2.8  &14.5 &0.56  &0.3  &0.23 &0.6   & 0\\
\end{tabular}  \end{ruledtabular}
\end{table*}

\section{Collision Geometry and Jet Absorption: the Model}
\label{sec:model}

Our discussion is limited to centrality dependence of the
suppression pattern over the $p_T$ range of 4.5 to 10 GeV/$c$,
where the hadron suppression is approximately
constant~\cite{ppg023}. Throughout the discussion, the average
suppression values for each centrality bin from the data are used
to compare with our calculation. Furthermore, since hadron
production in this $p_T$ region seems to be consistent with vacuum
jet fragmentation, we neglect the medium modification of jet
fragmentation and assume that the suppression for partons is
identical to the one observed for hadrons\footnote{Due to the
steeply falling power-law spectra of scattered partons, the hadron
$p_T$ spectra is dominated by fragments biased towards large $z$.
As was shown in Ref.\cite{jacob}, the shape of the hadron spectra
becomes nearly identical to the shape of the parton spectra,
independent of the fragmentation process. Since we model jet
absorption at the parton level, the suppression of hadrons thus
should be approximately the same as the suppression of partons.}.

The collision geometry is modelled by a Monte Carlo simulation of
Au + Au collisions based on the Glauber approach~\cite{glauber1}.
For the Au nucleus a Woods-Saxon density distribution with
radius $R~=~6.38$ fm and diffusivity $a~=~0.53$ fm is
used~\cite{glauber2}. We calculate for each simulated Au + Au
collision the underlying number of nucleon-nucleon collisions
($N_{\textrm{coll}}$) and the number of participating nucleons ($N_{\textrm{part}}$),
assuming the inelastic nucleon-nucleon cross section is
$\sigma^{\textrm{inel}}_{NN} = 42$ mb.

Centrality classes are defined from the fractional cross section
by cutting on the impact parameter of the collisions. However, the
centrality determination is insensitive to the specific cuts and
consequently methods employed
elsewhere~\cite{ppg001,star,dimacent} give similar results. The
average number of participants, nucleon-nucleon collisions and
impact parameter for different centrality classes are listed in
Table.~\ref{tab:model1}. For each centrality class, we also
calculate the participant density profile, $\rho_{\textrm{part}}(x,y)$, and
the nucleon-nucleon collision density profile, $\rho_{\textrm{coll}}(x,y)$,
in the plane transverse to the beam direction. The density
profiles for a central event class and a peripheral class are
shown in Fig.~\ref{fig:mden}; the peak values of the collision and
participant densities are also listed in Table~\ref{tab:model1}
for all centrality classes.

In the following we assume that the energy density is proportional
to the participant density. This is motivated by the recognition
that the bulk particle production scales approximately with the
number of participants~\cite{npartscale,sasha,Bialas,dimacent}. In
a later section we study the sensitivity of our results on this
assumption. To give the participant density a physical scale we
relate it to the energy density using the Bjorken
estimate~\cite{bjorken},
\begin{equation}
\epsilon_{\textrm{bj}} = \frac{E}{\tau_0A}\propto
\frac{N_{\textrm{part}}}{\tau_0\pi
r_0^2\left(\frac{N_{\textrm{part}}}{2}\right)^{2/3}}
\end{equation}
where $\tau_0$ = 0.2 -- 1 $\textrm{fm}/c$ is the formation time, $r_0$ = 1.2
fm is the effective nucleon radius, and $N_{\textrm{part}}$ is the number
of participating nucleons. For central collisions the
experimentally determined value of $\epsilon_{\textrm{bj}}$ is $\approx$ 5
GeV/$\textrm{fm}^3$~\cite{ppg002} for $\tau_0$~=~1~$\textrm{fm}/c$. With approximately
350 participants the scale factor to convert participant density
to $\epsilon_{\textrm{bj}}$ is 2 GeV/$\textrm{fm}$ for $\tau_0 = 1~\textrm{fm}/c$ or 10
GeV/$\textrm{fm}$ for $\tau_0 = 0.2~\textrm{fm}/c$.

Binary scaling of hard scattering assumes that the incoming parton
distribution in Au + Au collisions is a superposition of the
individual nucleon parton distribution functions. According to the
factorization theorem the probability for the hard scattering
processes in Au + Au is then proportional to $\rho_{\textrm{coll}}$.
Therefore we generate back-to-back parton pairs in the transverse
plane with a distribution of $\rho_{\textrm{coll}}(x,y)$ and isotropically
in azimuth. These partons are then propagated through the nuclear
medium with density proportional to $\rho_{\textrm{part}}(x,y)$. The
survival probability of a parton produced at $(x,y)$ along
direction ($n_{x}$,$n_{y}$) is calculated as~\cite{shuryak1}
\begin{equation}
f = \textrm{exp}(- \kappa I) \qquad ,
\label{eq:1}
\end{equation}
where $\kappa$ is the absorption strength, which is the only free
parameter in the model. $I$ is the matter integral along the path
of the parton, which is calculated as,
\begin{equation}
    I=\int_{0}^\infty dl\hspace{2mm} l\frac{l_0}{l+l_0} \rho{\left(x+\left(l+l_0\right)
n_x,y+\left(l+l_0\right)n_y\right)} \label{eq:2}
\end{equation}
This parameterization corresponds to a quadratic dependence of the
absorption ($\propto l dl$) in a longitudinally expanding medium
($\frac{l_0}{l+l_0}$)~\footnote{The effect due to transverse
expansion was shown to be small~\cite{vitevv2}.} . Here we assume
that partons travel with speed of the light and that they sense
the dense matter after a formation time of 0.2 $\textrm{fm}/c$ or a
distance of $l_0$ = 0.2 fm. However, the results presented in
this paper do not depend strongly on the choice of $l_0$.

The absorption strength $\kappa$ is then fixed in central Au + Au
collisions to reproduce the observed hadron suppression. In the
following, we use $\kappa$ = 0.7 which gives a suppression factor
of 4.35 as measured for 0--5\% most central collisions by
PHENIX~\cite{ppg023}. This corresponds to an absorption length of
$\lambda\sim3.4$ fm for a parton traversing an expanding medium
with a constant participant density of 2.48 $\textrm{fm}^{-2}$.

\begin{figure}
\begin{center}
\resizebox{0.6\columnwidth}{!}{\includegraphics{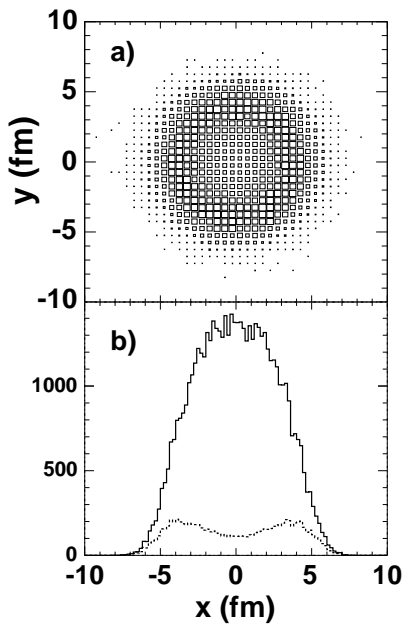}}
\caption{\label{fig:profpart} a) Origination point distribution in
the transverse plane for partons escaping the dense medium created
in central collisions with $\langle b\rangle\sim 2.2$~fm. b)
Generated (solid line) and survived (dotted line) parton initial
$x$ position for the cut $|y| < 1$~fm.}
\end{center}
\end{figure}

Fig.~\ref{fig:profpart} shows the origination position $(x,y)$ of
the hard-scattering partons that escape the medium in 0--5\% most
central collisions. The depletion at the center of the overlap
region is evident. The origins of the surviving jets are almost
uniformly distributed over the collision region, which biases jet
emission towards the surface of the collision region. Partons
emitted from the center of the collision region typically traverse
matter with density of more than one participant per fm$^2$ for a
distance of $\gtrsim5$ fm and thus are frequently absorbed,
while partons generated near the surface can easily escape if
emitted towards the surface. Therefore, the short absorption
length (less than half of the radius of the nuclear overlap)
naturally leads to emission of jets and thus of high $p_T$ hadrons
from outer layer of the medium.

\section{Centrality dependence of jet absorption}

\begin{figure*}[ht]
\begin{center}
\includegraphics[width=0.85\linewidth]{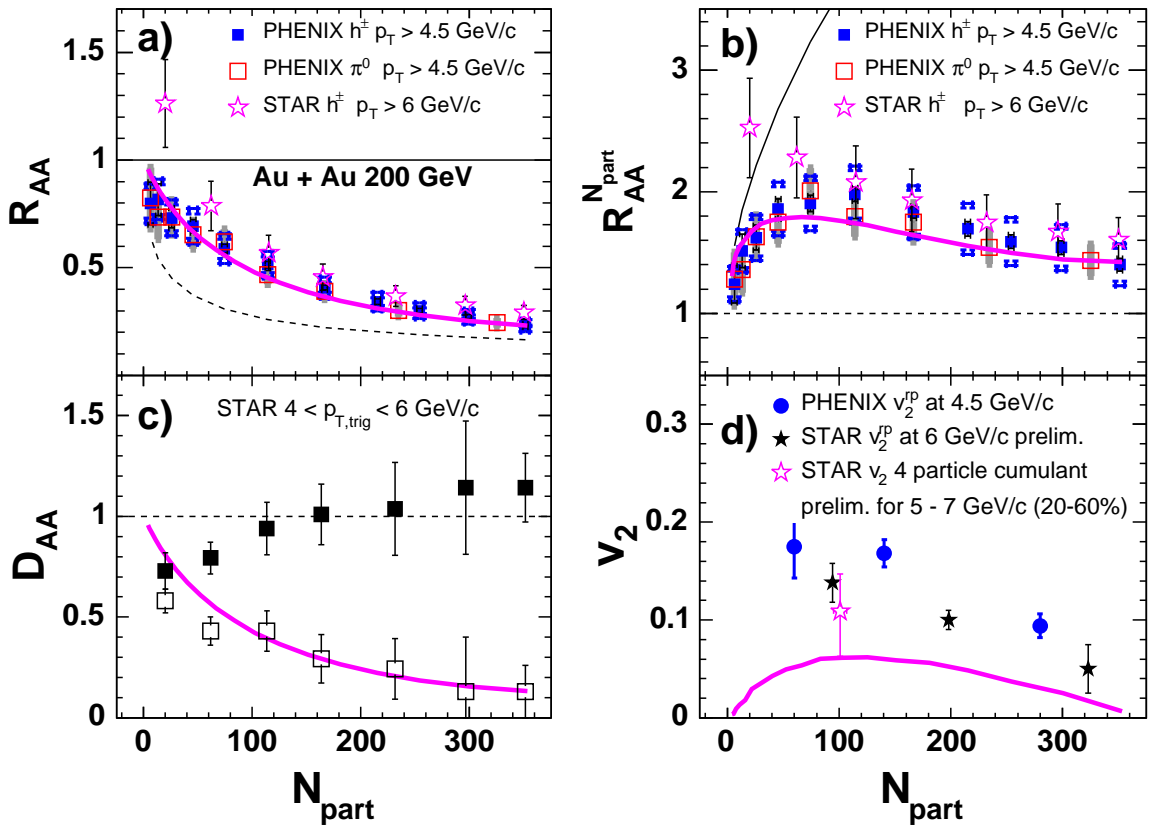}
\caption{\label{fig:modelws1} (Color online) Centrality dependence
of normalized yield of high $p_T$ hadrons ( a and b ),
back-to-back correlation strength (c), and $v_2$ (d) at high
$p_T$. In all four panels, the thick solid line indicates the
result of our calculations based on Eqs.~\ref{eq:1}-\ref{eq:2}.
Further details are discussed in the text. }
\end{center}
\end{figure*}

Once $\kappa$ has been determined, the survival rate of jets, the
probability to find back-to-back jets, and also the azimuthal
anisotropy of emitted jets relative to the reaction plane is fixed
for any centrality class. In Fig.~\ref{fig:modelws1} we compare
our model calculation to the centrality dependence of three
compilations of experimental data:

\begin{enumerate}

\item
Charged hadron and $\pi^0$ data from PHENIX~\cite{ppg014,ppg023}
and charged hadron data from STAR~\cite{starch} are plotted on the
top panels of Fig.~\ref{fig:modelws1}. Shown are the ratios of the
observed high $p_T$ hadron yields relative to the expected yield
from the underlying nucleon-nucleon collisions, normalized to
$N_{\textrm{coll}}$, $R_{AA}$ in panel a), or $N_{\textrm{part}}$/2, $R_{AA}^{N_{\textrm{part}}}$ in panel b),
of the specific centrality class. The PHENIX data have a lower
$p_T$ cut of 4.5 GeV/$c$, while the STAR data are given for
$p_T >$~6 GeV/$c$.

\item
The bottom left panel (c) gives the back-to-back jet correlation
strength measured by STAR~\cite{starbtob}. The away side
correlation strength can be defined by~\cite{wangnew}
\\
\\
\begin{eqnarray}
&&D_{AA}(p_T^{\textrm{trig}}) =\\\nonumber
&&\int^{p_T^{\textrm{trig}}}_{p_{0}}dp_T
\int^{|\phi_1-\phi_2|>\phi_0}d\phi
\frac{d\sigma_{AA}^{h_1h_2}/d^2p_T^{\textrm{trig}}dp_T
d\phi}{d\sigma_{AA}^{h_1}/d^2p_T^{\textrm{trig}}}
\end{eqnarray}
for an associated hadron $h_2$ with $p_T$ in backward azimuthal
direction of a trigger hadron $h_1$ with $p_T^{\textrm{trig}}$. The STAR
data are for charged trigger particles with $4 < p_T^{\textrm{trig}} < 6$
GeV/$c$ and associated charged hadrons with $p_T > p_{0} = 2$
GeV/$c$ detected within $|\phi_1 - \phi_2| > \phi_0 = 2.24$. The data
are normalized to the expectation from $p$ + $p$ collisions, and
corrected for combinatorial random background and the azimuthal
anisotropy of bulk particle production.

\item
The azimuthal anisotropy of high $p_T$ charged particles,
quantified by $v_2$, the second Fourier coefficient of the
$dN/d\phi$ distribution, is shown in the bottom right panel (d).
The PHENIX data are measured for $4~<~p_T~<~5$ GeV/$c$ with respect to
the reaction plane~\cite{ppg022}. The preliminary $v_2$ values
from STAR experiment were determined using reaction plane
method~\cite{filimonov} around 6 GeV/$c$ and 4 particle cumulants
for $5~<~p_T~<~7$ GeV/$c$~\cite{snellings}.
\end{enumerate}

The results of our model calculation are shown as thick solid
lines on all panels. Once the absorption strength is adjusted to
the observed yield from the most central collisions, the
centrality dependence of the normalized yield is well reproduced.
In Fig.~\ref{fig:modelws1}b, the calculated $R_{AA}^{N_{\textrm{part}}}$ increases for
peripheral collision, which is expected if the yield scales with
the $N_{\textrm{coll}}$ (thin solid line). Only a small fraction of the
partons is absorbed since the matter density and volume are small.
As the centrality increases both matter density and volume
increase; for collisions with more than 100 participants
absorption overwhelms the increase due to point like scaling and
the $R_{AA}^{N_{\textrm{part}}}$ decreases with centrality.

The jet absorption model also reproduces the magnitude and
centrality dependence of the back-to-back correlations without
further adjustments. The calculated suppression is negligible for
peripheral collisions, increases continuously and reaches a
suppression factor of 7 for the most central bin, consistent with
the data within errors\footnote{We notice, however, due to the
large errors of the STAR data, the suppress factor for central
collisions is not well constrained.}. The model suggests that the
suppression of back-to-back correlations is almost a factor of 1.6
stronger than that for the single inclusive hadron yield. In the
jet absorption model, most of the surviving partons come from near
the surface of the overlap region. Thus their partner partons have
to traverse on average a longer distance through the medium. This
naturally leads to a stronger suppression of the back-to-back
correlation. It should be noted that in our model, we assume that
the jet fragments outside of the medium, in other words, the
fragmentation is identical to $p$ + $p$ by construction. Therefore,
the near angle jet correlation strength can not deviate from unity
in our model.

Due to the asymmetry of the overlap region of the two nuclei, the
average amount of matter traversed by a parton depends on its
azimuthal direction with respect to the reaction plane, which
leads to an azimuthal anisotropy of the emitted
jets~\cite{xnwang3}. This anisotropy reaches its maximum for
collisions with an impact parameter of about 9 fm corresponding
to approximately 100 participants. It is small for peripheral and
central collisions. Our calculation reproduces the measured trend
of the centrality dependence of $v_2$, but the magnitude is below
the measured value~\footnote{We note that the 6\% $v_2$ from our
calculations corresponds to a factor of 1.3 larger jet survival
probability in reaction plane than out of plane.}. With a matter
density profile deduced from a Woods-Saxon distribution, a large
fraction of the surviving jets is emitted from the low density
region at large radii (see Fig.~\ref{fig:profpart}b). Therefore
the anisotropy is diluted. Larger values of $v_2$ are obtained for
larger $\kappa$, but these reproduce neither the suppression of
the yield nor the back-to-back correlation strength. For a
Woods-Saxon nuclear profile, $v_2$ reaches a maximum at a certain
$\kappa$ (for $b$ = 9 fm, $v_2^{\textrm{max}}$ = 10\%), but then decrease
to 0 as $\kappa\to\infty$ because all surviving jets come from the
diffuse outer layer of the overlap region, which is essentially
isotropic~\cite{shuryak2}. The calculated $v_2$ values are very
sensitive to the actual nuclear profile used, we shall come back
to this in Section~\ref{sec:denprof}.

\section{Discussions}
\subsection{Dependence on Absorption Pattern}
\label{sec:abs}

Parton energy loss through gluon bremsstrahlung is thought to be
proportional to the path length squared in a static
medium~\cite{BDMPS,baier}. This motivated our ansatz in
Eq.~\ref{eq:2}. We note that although $I$ calculated from
Eq.~\ref{eq:2} was interpreted as absorption $\propto l^2$ in a
longitudinally expanding medium it can also be treated as
absorption $\propto l$ in a static medium when $l_0$ is small. In
the following, we denote it as $I_1$. In order to test the
sensitivity of the data to discriminate between different types of
energy loss we repeat the calculation for two additional types of
absorption,
\begin{eqnarray}
\label{eq:3}
    I_2&=&\int_{0}^\infty dl \hspace{2mm}\frac{l_0}{l+l_0}\rho{\left(x+\left(l+l_0\right)n_x,y+\left(l+l_0\right)n_y\right)}\nonumber\\
    I_3&=&\int_{0}^\infty dl \hspace{2mm}l\rho{\left(x+ln_x,y+ln_y\right)}
\end{eqnarray}
$I_2$ assumes absorption proportional to the path length $l$ in a
longitudinally expanding medium as one may expect for a hadronic
energy loss~\cite{gallmeister} scenario; $I_3$ assumes a static
source and absorption $\propto l^2$. Again the absorption strength
is adjusted to give a factor of 4.35 suppression of the yield for
central collisions. We find $\kappa = 0.84$~fm ($\lambda = 1.9$~fm) and
$\kappa = 0.06$ ($\lambda = 3.6$~fm) for $I_2$ and $I_3$, respectively.

\begin{figure}[ht]
\begin{center}
\begin{tabular}{c}
\resizebox{\columnwidth}{!}{\includegraphics{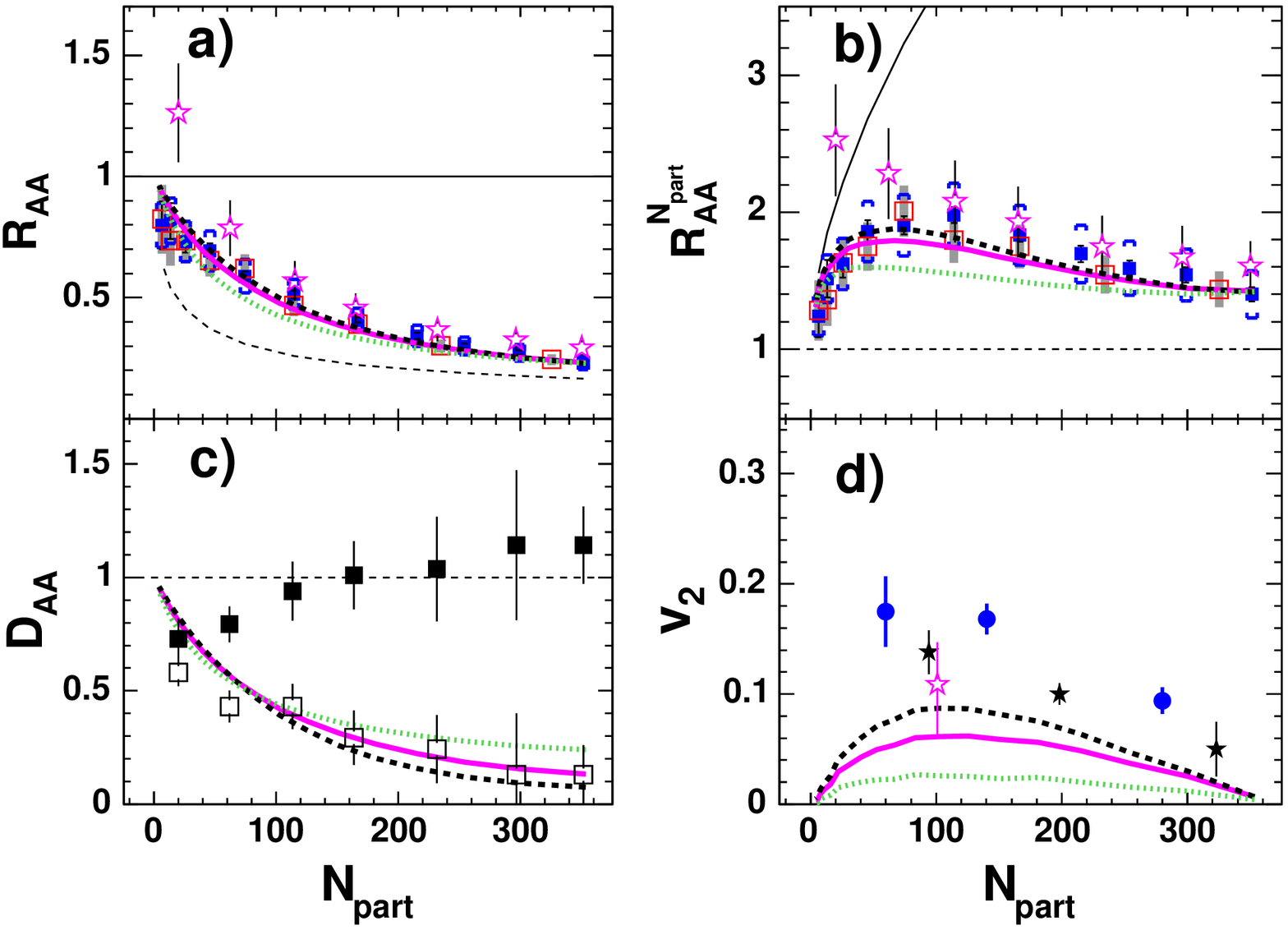}}
\end{tabular}
\caption{\label{fig:abs} (Color online) Calculation of $R_{AA}$ (a)
$R_{AA}^{N_{\textrm{part}}}$ (b), $D_{AA}$ (c), and $v_2$ (d) using a Woods-Saxon
nuclear profile for $I_1$ (solid line), $I_2$ (dotted line), and
$I_3$ (dashed line) types of jet absorption.}
\end{center}
\end{figure}

The results are compared with each other in Fig.~\ref{fig:abs}.
The centrality dependence of the normalized yield is similar for
the three different absorption patterns and thus gives little
discrimination power. The correlation strength $D_{AA}$ is more
sensitive, but the largest sensitivity is found for the anisotropy
parameter $v_2$. The $I_2$ scenario tends to localize the
absorption in the region within the distance of $\approx l_0$ from
the jet creation point. In this case, the suppression is dominated
by the initial matter profile and is insensitive to the later
evolution of the system. This naturally leads to a similar value
between $R_{AA}$ and $D_{AA}$. The surviving jets are also emitted
more isotropically resulting in a smaller $v_2$~\footnote{In this
analysis, we have ignored the transverse expansion, which can
further reduce $v_2$, but only slightly change $R_{AA}$ and
$D_{AA}$~\cite{vitevv2}.}.

On the other hand, the absorption in a static medium with
quadratic path length dependence ($I_3$) has a strong dependence
on the jet path. Partons are absorbed along their full trajectory
in the medium. Thus both the suppression of the $D_{AA}$ and the
value of $v_2$ depend more strongly on the centrality of the
collision. In particular, one observes a stronger centrality
dependence of $D_{AA}$ than of the $R_{AA}$ and a larger $v_2$.
Although the magnitude of $v_2$ is still below the experimental
value.

\subsection{Dependence on Density Profile}
\label{sec:denprof} Motivated by the approximate $N_{\textrm{part}}$ scaling
of bulk particle production, we have assumed that the initial
energy density created is proportional to the participant density
($\epsilon \propto \rho_{\textrm{part}}$). A closer look at the data
reveals that both the transverse energy $E_T$ and particle
multiplicity increase faster than linear with
$N_{\textrm{part}}$~\cite{ppg001,sasha,phobos2}. In a two component
model~\cite{twocomponent}, the additional increase is attributed
to (mini)jet production, which should scale with $N_{\textrm{coll}}$. To test
the sensitivity of our model to the initial energy density
profile, we repeated the calculation assuming that the energy
density is proportional to the collision density
$\rho_{\textrm{coll}}(x,y)$.

\begin{figure}[ht]
\begin{center}
\begin{tabular}{c}
\resizebox{\columnwidth}{!}{\includegraphics{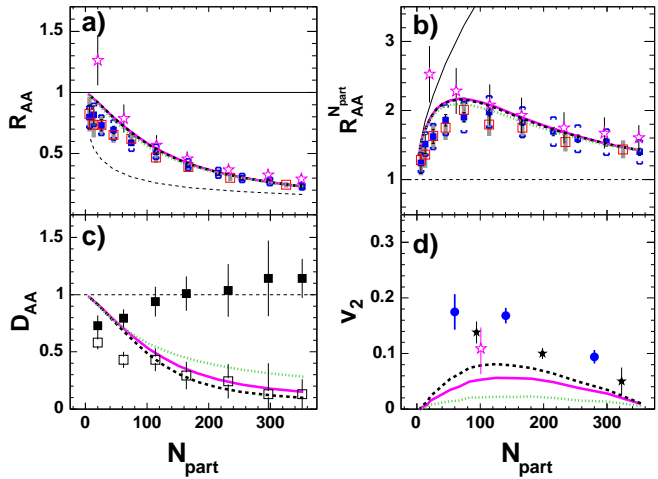}}
\end{tabular}
\end{center}
\caption{\label{fig:density} (Color online) Calculation of $R_{AA}$
(a), $R_{AA}^{N_{\textrm{part}}}$ (b), jet correlation (c), and $v_2$ (d) assuming
$\epsilon \sim \rho_{\textrm{coll}}$. The three curves are for $I_1$ (solid
line), $I_2$ (dotted line), and $I_3$ (dashed line) types of jet
absorption, respectively.}
\end{figure}

The results of the calculation are presented in
Fig.~\ref{fig:density}. Generally the agreements between the data
and the model calculation are equally good for both matter density
profiles. In detail the suppression of the inclusive yield
(Fig.~\ref{fig:density}a and b) shows a slightly stronger
centrality dependence. This is easily explained: the absorption
strength $\kappa$ is fixed to reproduce the suppression of the
yield in the most central collisions;
$\left<\rho_{\textrm{coll}}(x,y)\right>$ varies by factor of 4 faster than
$\left<\rho_{\textrm{part}}(x,y)\right>$ from peripheral to central
collisions. As a result, the absorption is smaller in peripheral
collisions. Similarly $D_{AA}$ is less suppressed in peripheral
collisions, but decrease faster for intermediate centralities. On
the other hand, the centrality dependence and magnitude of $v_2$
are similar to Fig.~\ref{fig:abs}, indicating that the two density
profiles produce an almost identical anisotropy for all three
absorption patterns.

In some recent model calculations nuclear distributions different
from a Woods-Saxon profile have been used, typically either a
hard-sphere~\cite{wangnew,shuryak1} or cylindrical nuclear
distributions~\cite{dima,nonflow,mueller}. The results are
significantly different for these unrealistic approximations. In
particular, $v_2$ is increased, where $v_2^{\textrm{cylindrical}}
> v_2^{\textrm{sphere}} > v_2^{\textrm{Woods-Saxon}}$.

\begin{figure*}[ht]
\begin{center}
\resizebox{1\linewidth}{!}{\includegraphics{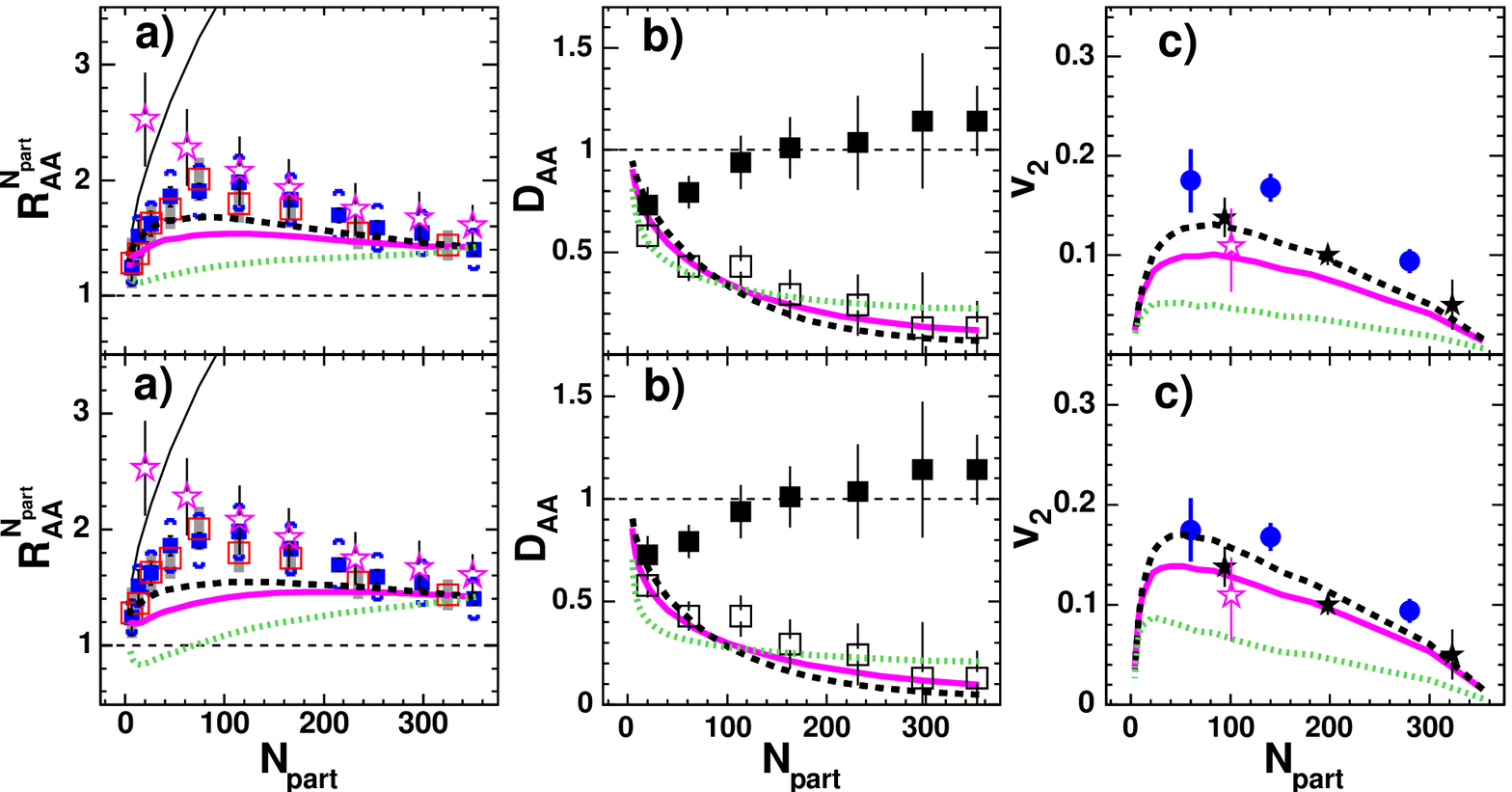}}\\
\caption{\label{fig:modelprof} (Color online) Calculation of
$R_{AA}^{N_{\textrm{part}}}$ (a), $D_{AA}$ (b), and $v_2$ (c) for a hard-sphere nuclear
profile (top panels) and a cylindrical nuclear profile (bottom
panels) using the $I_1$ (solid line), $I_2$ (dotted line), and
$I_3$ (dashed line) form of jet absorption.}
\end{center}
\end{figure*}

Fig.~\ref{fig:modelprof} shows the calculations for a hard-sphere
nuclear distribution in the top panels and for a cylindrical
nuclear distribution in the bottom panels. In both cases, all
three absorption scenarios miss the centrality dependence of the
normalized inclusive yield. In contrast to the data, the
suppression sets in at rather peripheral collisions and remain
approximately constant with centrality.

On the other hand, cylindrical and spherical profiles apparently
result in much better agreement with $v_2$ data. The reason is
that both cylindrical and hard-sphere collision profiles have
sharp surfaces with a large eccentricity, leading to a larger
$v_2$~\cite{v2max}. In contrast, a Woods-Saxon collision profile
has a diffuse surface. While it may be conceivable that the
density profile of the matter produced in the collision deviates
from the convolution of two Woods-Saxon nuclear distributions, it
is hard to imagine that the probability for hard scattering
deviates from a Woods-Saxon density distribution significantly.

\subsection{Energy and System Size Dependence}

If jet suppression is really dominated by final state energy loss,
it should decrease if the energy density or the volume of the
medium is reduced. In addition to varying the impact parameter, both
volume and density can be reduced by varying the mass number of
the colliding nuclei. Alternatively the density may also be
changed by varying the beam energy $\sqrt{s_{NN}}$.

The jet absorption model can provide a simple baseline prediction
for both the beam energy and system size dependence of the
suppression, if we assume that the absorption strength $\kappa$ is
the same as in $\sqrt{s_{NN}}$=200 GeV Au + Au collisions. We limit the
discussion to the calculation presented in
section~\ref{sec:model}.

\begin{figure}[ht]
\begin{center}
\begin{tabular}{c}
\includegraphics[scale =0.4]{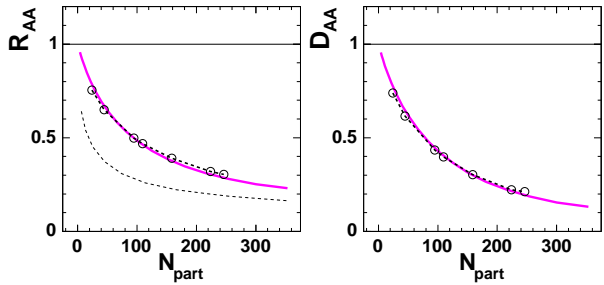}
\end{tabular}
\end{center}
\caption{\label{fig:spescan} (Color online) The calculated system
size dependence (open circles) of the $R_{AA}$ (left) and $D_{AA}$
(right) compared with their centrality dependence in Au + Au
collisions (thick solid line). The open circles are calculated for
0--5\% most central collisions for each collision system. The thick
solid line represent the calculation for $I_1$ type of absorption
(same calculation shown in Fig.~\ref{fig:modelws1}b and c.)}
\end{figure}

\begin{figure}[ht]
\resizebox{1\columnwidth}{!}{\includegraphics{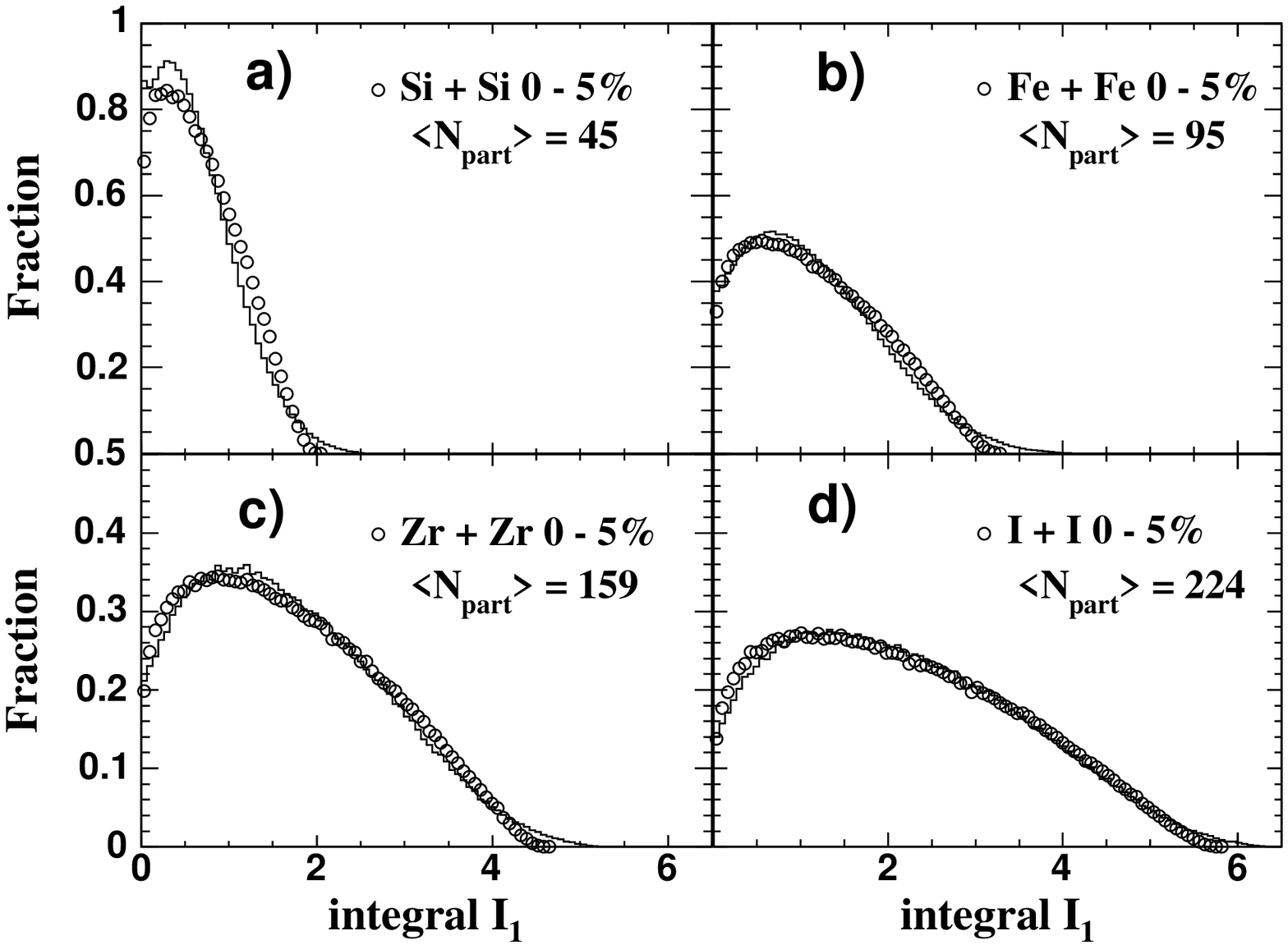}}
\caption{\label{fig:inteabs}The matter integral $I_1$
distributions as defined in Eq.~\ref{eq:2} for central a) Si + Si,
b) Fe + Fe, c) Zr + Zr, d) I + I collisions (solid circles)
compared with Au + Au collisions (solid lines) at different impact
parameters with similar $\langle N_{\textrm{part}} \rangle$.}
\end{figure}

The system size dependence can be evaluated without further
assumptions. In Fig.~\ref{fig:spescan} the nuclear modification
factor $R_{AA}$ for central collisions of smaller systems (see
Table.~\ref{tab:system}) at $\sqrt{s_{NN}}$=200 GeV is compared to the
calculated centrality dependence for Au + Au collisions (see
Fig.~\ref{fig:modelws1}). The agreement of the absolute values and
the $N_{\textrm{part}}$ dependence of the $R_{AA}$ and $D_{AA}$ is
remarkable. This implies that to the first order the volume is
proportional to $N_{\textrm{part}}$ and that the participant densities
profiles are very similar. This similarity is illustrated in
Fig.~\ref{fig:inteabs} by comparing the integral distribution
$I_1$ experienced by the generated partons in central collisions
for various collision systems with that for centrality selected Au
+ Au collisions with similar $N_{\textrm{part}}$. Since the jet correlation
strength also depends on the volume and average density, the
calculated $D_{AA}$ is also similar for light systems and
corresponding Au + Au collisions. In contrast, the anisotropy
calculated for central collisions of smaller nuclei is consistent
with zero as expected.
\begin{table*}
\caption{\label{tab:system} List of light collisions systems and
corresponding $N_{\textrm{part}}$ for 0--5\% central collisions.}
\begin{ruledtabular} \begin{tabular}{lllllllll}
Species&O + O& Si + Si& Fe + Fe& Cu + Cu& Zr + Zr& I + I& La + La\\\hline
$N_{\textrm{part}}$&24&45&95&109&159&224&246\\
\end{tabular}   \end{ruledtabular}
\end{table*}

To guide our estimate of the beam energy dependence we assume that
the matter density scales like the Bjorken energy density. For an
average mix of quark and gluon jets which we do not vary with
$\sqrt{s}$, we fix the absorption coefficient $\kappa$ to be the
same as in $\sqrt{s_{NN}} = 200$ GeV. In this estimation the energy density
increases by a factor of 2 from SPS ($\sqrt{s_{NN}}$ = 17 GeV) to RHIC
($\sqrt{s_{NN}}$ = 200 GeV) energies. Interpolating between these values
and scaling down the density in our calculation accordingly, we
find that the high $p_T$ hadron suppression is still a factor
$\sim$3 for lower energy RHIC run at $\sqrt{s_{NN}}$ = 62.4 GeV.

Our estimation would also predict a factor of 2 suppression at SPS
energies, which is not observed experimentally. However, data at
the SPS~\cite{SPS} are limited to $p_T$ below 4 GeV/$c$, a region
where other mechanism like $p_T$ broadening and hydrodynamic flow
complicate the interpretation of particle production in heavy ion
collisions. We noted that the predictive power of our approach is
limited, because we did not take into account the $\sqrt{s_{NN}}$
dependence of the quark (gluon)jet fraction, the hard-scattering
cross section and the system lifetime, which could be important.

\section{Conclusion}
We have shown that the experimentally observed centrality
dependence of the suppression of the hadron yield $R_{AA}$ and the
suppression of back-to-back correlation $D_{AA}$ can be
quantitatively described by jet absorption in an opaque medium
when a realistic nuclear geometry is included. In our model, the
centrality dependence of $R_{AA}$ and of the $D_{AA}$ are rather
insensitive to the details of the medium density profile. $R_{AA}$
is also insensitive to the absorption patterns. Both $R_{AA}$ and
$D_{AA}$ do not distinguish between central collisions of small
systems and centrality selected Au + Au collisions with similar
$N_{\textrm{part}}$. Our interpretation is that $R_{AA}$ and $D_{AA}$ depend
mostly on volume and average density of the opaque medium. In the
limit of very opaque matter, $R_{AA}$ probably will not reveal
further insight into the details of the absorption pattern or
alternatively of the energy loss and the density profile, unless
data at much higher $p_T$ become available.

We find that observables like $D_{AA}$ and $v_2$ are more
sensitive to the absorption patterns. Our model does not describe
$v_2$ quantitatively unless an unrealistic nuclear density profile
is used. This might indicate that the actual diffuseness of the
opaque medium is smaller than expected from the density profile
obtained by convoluting two Woods-Saxon nuclear
distributions~\cite{xnwang3}. It is also conceivable that the real
suppression is larger, and that the observed suppression is
reduced by ``soft`` particles from dynamic mechanisms different
from jet fragmentation, such as hydrodynamics~\cite{hydro} plus
viscosity correction~\cite{derek}, quark
coalescence~\cite{coalence}, and quark/diquark pick
up~\cite{shuryak2}. In this case, both the ``soft`` particles and
a stronger suppression would lead to a larger $v_2$.

{\bf Acknowledgments:} We greatly appreciate the fruitful
discussions with P. Kolb, E. Shuryak, and J.C. Solana. This work
was supported by the Department of Energy, Office of Science, and
Nuclear Physics Division under Grant No. DE-FG02-96ER40988.


\begin{thebibliography}{99}
\bibitem{Gyu90} M.~Gyulassy and M.~Pl\"umer, \Journal{\PLB}{243}{432}{1990};
X.N.~Wang and M.~Gyulassy, \Journal{\PRL}{68}{1480}{1992};
R.~Baier {\it et al.}, \Journal{\PLB}{345}{277}{1995}.
\bibitem{BDMPS} R.~Baier, D.~Schiff and B.G.~Zakharov, Annu. Rev. Nucl. Part. Sci. {\bf 50}, 37 (2000).
\bibitem{baier} R.~Baier, Y.L.~Dokshitzer, A.H.~Mueller and D.~Schiff,
J. High Energy Phys. {\bf 09}, 033 (2001)
\bibitem{ppg014} S.S~Adler {\it et al.}, [PHENIX Collaboration], \Journal{\PRL}{91}{072301}{2003}.
\bibitem{ppg023} S.S~Adler {\it et al.}, [PHENIX Collaboration], \Journal{\PRC}{69}{034910}{2004}.
\bibitem{starch} J.~Adams {\it et al.}, [STAR Collaboration], \Journal{\PRL}{91}{172302}{2003}.
\bibitem{starbtob} C.~Adler {\it et al.}, [STAR Collaboration], \Journal{\PRL}{90}{082302}{2002}.
\bibitem{dauphobos} B.B.~Back {\it et al.}, [PHOBOS Collaboration], \Journal{\PRL}{91}{072302}{2003}.
\bibitem{ppg028} S.S~Adler {\it et al.}, [PHENIX Collaboration], \Journal{\PRL}{91}{072303}{2003}.
\bibitem{daustar} J.~Adams {\it et al.}, [STAR Collaboration], \Journal{\PRL}{91}{072304}{2003}.
\bibitem{daubrahms} I.~Arsene {\it et al.}, [BRAHMS Collaboration], \Journal{\PRL}{91}{072305}{2003}.
\bibitem{starks} J.~Adams {\it et al.}, [STAR Collaboration],
\Journal{\PRL}{92}{052302}{2004}.
\bibitem{ppg015} S.S~Adler {\it et al.}, [PHENIX Collaboration], \Journal{\PRL}{91}{172301}{2003}.
\bibitem{bjorken2} J.D.~Bjorken, FERMILAB-PUB-82-059-THY  (unpublished).
\bibitem{xinnianprl} E.~Wang and X.N.~Wang,
\Journal{\PRL}{87}{142301}{2001}.
\bibitem{ivanprl} M.~Gyulassy,  P.~Levai and I.~Vitev, \Journal{\PRL}{85}{5535}{2000}.
\bibitem{ursprl} C.A.~Salgado and U.A.~Wiedemann, \Journal{\PRL}{89}{092303}{2002}.
\bibitem{baier2} R.~Baier, Yu.L.~Dokshitzer, A.H.~Mueller and D.~Schiff, \Journal{\NPB}{531}{403}{1998}.
\bibitem{xnwang3} M.~Gyulassy, I.~Vitev and X.N.~Wang, \Journal{\PRL}{86}{2537}{2001}.
\bibitem{vitevv2} M.~Gyulassy, I.~Vitev, X.N.~Wang and
P.~Huovinen, \Journal{\PLB}{526}{301}{2002}.
\bibitem{mueller} B.~M\"uller, \Journal{\PRC}{67}{061901}{2003}.
\bibitem{nara} T.~Hirano and Y.~Nara, \Journal{\PRL}{91}{082301}{2003}.
\bibitem{wangnew} X.N.~Wang, \Journal{\PLB}{595}{165}{2004}.
\bibitem{shuryak2} J.C.~Solana and E.V.~Shuryak, {\tt hep-ph/0305160}.
\bibitem{shuryak1} E.V.~Shuryak, \PRC{66} 027902 (2002).
\bibitem{jacob} M.~Jacob and P.V.~Landshoff, \Journal{\PR}{48}{285}{1978}.
\bibitem{glauber1} R.J.~Glauber and G.~Matthiae, Nucl.\ Phys.\ B {\bf 21}, 135 (1970).
\bibitem{glauber2} B.~Hahn, D.G.~Ravenhall and R.~Hoftstadter, \Journal{\PR}{101}{1131}{1956}.
\bibitem{ppg001} K.~Adcox {\it et al.}, [PHENIX Collaboration], \Journal{\PRL} {86}{3500}{2001}.
\bibitem{dimacent} D.~Kharzeev and M.~Nardi, \Journal{\PLB}{507}{121}{2001}.
\bibitem{star} C.~Adler {\it et al.}, [STAR Collaboration],
\Journal{\PRL}{89}{202301}{2002}.
\bibitem{Bialas} A.~Bialas, A.~Bleszynski and W.~Czyz, \Journal{\NPB}{111}{461}{1976}.
\bibitem{sasha} A.~Bazilevsky, for the PHENIX Collaboration, \Journal{\NPA}
{715}{486}{2003}.
\bibitem{npartscale}
NA49 Collaboration, P.G.~Jones {\it et al.},
\Journal{\NPA}{610}{188}{1996}; WA98
Collaboration, T.~Peitzmann {\it et al.}, \Journal{\NPA}{610}{200}{1996}; WA80 Collaboration, R.~Albrecht {\it
et al.}, \Journal{\PRC}{44}{2736}{1998}.
\bibitem{bjorken} J.D.~Bjorken, \Journal{\PRD}{27}{140}{1983}.
\bibitem{ppg002} K.~Adcox {\it et al.}, [PHENIX Collaboration], \Journal{\PRL} {87}{052301}{2001}.
\bibitem{ppg022} S.S~Adler {\it et al.}, [PHENIX Collaboration], \Journal{\PRL}{91}{182301}{2003} .
\bibitem{filimonov} K.~Filimonov, \Journal{\NPA}{715}{737}{2003}.
\bibitem{snellings} R.~Snellings, [STAR Collaboration],
Heavy Ion Phys.\  {\bf 21}, 237 (2004).
\bibitem{gallmeister} K.~Gallmeister, C.~Greiner and Z.~Xu, \Journal{\PRC}{67}{044905}{2003}.
\bibitem{phobos2} B.B.~Back {\it et al.}, [PHOBOS Collaboration], \Journal{\PRC}{67}{021901R} {2003}.
\bibitem{twocomponent} S.~Li and X.N.~Wang \Journal{\PLB}{527}{85}{2002}.
\bibitem{dima} D.~Kharzeev, E.~Levin and L.~McLerran, \Journal{\PLB}{561}{93}{2003} and references therein.
\bibitem{nonflow} Y.V.~Kovchegov and K.L.~Tuchin,
\Journal{\NPA}{708}{413}{2002}.
\bibitem{v2max} S.A.~Voloshin, \Journal{\NPA}{715}{379}{2003}.
\bibitem{SPS} WA98 Collaboration, M.M.~Aggarwal {\it et al.},
\Journal{\EPJ}{23}{225}{2002}; CERES
Collaboration, G.~Agakishiev {\it et al.}, hep-ex/0003012; NA49 Collaboration, H.~Appleshauser, {\it et al.},
 \Journal{\PRL}{82}{2471}{1999}.
\bibitem{coalence} R.C.~Hwa and C.B.~Yang, \Journal{\PRC}{67}{034902}{2003};
R.J.~Fries, B.~Muller, C.~Nonaka and S.A.~Bass,
\Journal{\PRL}{90}{202303}{2003}.
\bibitem{hydro} P.F.~Kolb and U.~Heinz, review for 'Quark Gluon Plasma 3',
Editors: R.C.~Hwa and X.N.~Wang, World Scientific, Singapore,
{\tt nucl-th/0305084}.
\bibitem{derek} D.~Teaney, {\tt nucl-th/0301099}.
\end{thebibliography}
\end{document}